\documentclass[12pt]{article}
\usepackage{amsfonts,amssymb,amsmath}
\usepackage{epsfig}
\usepackage{enumerate}   

\newcommand\RR{\mathbb R}

\newcommand\beq{\begin{equation}}
\newcommand\eeq{\end{equation}}

\newtheorem{theorem}{Theorem}
\newtheorem{lemma}{Lemma}

\newtheorem{proposition}{Proposition}

\newtheorem{example}{Example}

\begin{document}

\title{Multipoint scatterers with \\ zero-energy bound states
\thanks{This work was fulfilled during the visit of the 
first author to the Centre de Math\'ematiques Appliqu\'ees of \'Ecole Polytechnique
in September-October 2016. The first author was also partially supported by by the program  
``Fundamental problems of nonlinear dynamics'' of the Presidium of RAS.}}
\author{P.G. Grinevich
\thanks{L.D. Landau Institute for Theoretical Physics RAS, Chernogolovka, Moscow region, Russia;
Moscow State University, Moscow, Russia; Moscow Physical-Technical Institute, 
Dolgoprudny, Russia; e-mail: pgg@landau.ac.ru} 
\and R.G. Novikov\thanks
{CNRS (UMR 7641), Centre de Math\'ematiques Appliqu\'ees, 
\'Ecole Polytechnique, Universit\'e Paris Saclay, 91128, Palaiseau, France;
IEPT RAS, 117997, Moscow, Russia; 
e-mail: novikov@cmap.polytechnique.fr}}
\date{}
\maketitle

\begin{abstract}
We study multipoint scatterers with zero-energy bound states in three dimensions. We present examples of such 
scatterers with multiple zero eigenvalue or with strong multipole localization of zero-energy bound states.
\end{abstract}

\section{Introduction}
We consider the model of point scatterers in three dimensions, which goes back to the classical works \cite{BP}, 
\cite{Fer}, \cite{Zel}, \cite{BF} and presented in detail in the book \cite{AGHH}. For more recent results 
on such models, see \cite{DFT}, \cite{BBMR}, \cite{GRN} and references therein. More precisely, we consider the 
stationary Schr\"odiger equation 
\beq
\label{eq:1}
-\Delta\psi + v(x)\psi = E\psi,  \ \ x\in\RR^3,
\eeq
with multipoint potential (scatterer) 
\beq
\label{eq:2}
v(x)=\sum\limits_{j=1}^{n} v_{z_j,\alpha_j}(x) ,
\eeq
consisting of $n$ single-point scatterers $v_{z_j,\alpha_j}(x)$, where each point scatterer $v_{z_j,\alpha_j}(x)$ is 
described by its position $z_j\in\RR^3$ and its internal parameter $\alpha_j\in\RR$, where $z_i\ne z_j$ if $i\ne j$.

In the present article we study multipoint scatterers $v$ for which equation (\ref{eq:1}) admits non-zero solutions 
$\psi\in L^2(\RR^3)$ at energy $E=0$, or in other words, we study the multipoint scatterers with zero-energy bound 
states. These studies are motivated, in particular, by studies of low-energy scattering effects in three dimensions. 
To our knowledge, the question about zero-energy bound states for multipoint scatterers was not considered properly 
in the literature. Besides, our studies were stimulated by \cite{TT}, where  interesting examples of regular rapidly 
decaying potentials with well-localized zero energy bound states in two dimensions were constructed using the 
Moutard transform technique. 

Results of the present article include Proposition~1, Theorem~1 and Examples~1 and 2 given below.

\section{Solitions of the Schr\"odinger equation with multipoint potential}
We say that $\psi$ satisfies (\ref{eq:1}) iff
\beq
\label{eq:3}
-\Delta\psi(x) = E\psi(x)  \ \ \mbox{for} \ \  x\in\RR^3\backslash\{z_1,z_2,\ldots,z_n\},
\eeq
and
\beq
\label{eq:4}
\psi(x) = \frac{\psi_{j,-1}}{|x-z_j|}+ \psi_{j,0}+O(|x-z_j|)  \ \ \mbox{as} \ \ x\rightarrow z_j, \ \ 
j=1,\ldots, n,
\eeq
where
\beq
\label{eq:5}
 \psi_{j,0}= 4\pi\alpha_j \psi_{j,-1}.
\eeq
In this article  we use the same normalization of multipoint scatterers as in the book \cite{AGHH}, see pages 47, 112. 

\begin{proposition} A function $\psi=\psi(x)$ satisfies (\ref{eq:3})-(\ref{eq:5}) if and only if this function admits the following representation:
\beq
\label{eq:6}
\psi(x)=\psi_0(x)+\sum\limits_{j=1}^{n} q_j G^+(|x-z_j|,E),
\eeq
where
\beq
\label{eq:7}
-\Delta\psi_0(x) = E\psi_0(x)  \ \ \mbox{for all} \ \  x\in\RR^3,
\eeq
\beq
\label{eq:8}
G^+(r,E)=-\frac{e^{i\sqrt{E}r}}{4\pi r}, \ \ r>0, \ \ i=\sqrt{-1}, \ \ \sqrt{E}\ge0 \ \ \mbox{for} \ \ E\ge0,
\eeq
and $\vec q=(q_1,\ldots, q_n)^t$ satisfies the following linear system:
\beq
\label{eq:9}
A \vec q=\vec\phi,
\eeq
where $A$ is the $n\times n$ matrix
\beq
\label{eq:10}
A_{j,j'}=\left\{\begin{array}{ll} 
                  \alpha_j -\frac{i\sqrt{E}}{4\pi} & \mbox{for} \ \ j'= j \\
                  G^+(|z_j-z_{j'}|,E) & \mbox{for} \ \ j'\ne j,
                \end{array}\right.
\eeq
\beq
\label{eq:11}
\vec\phi=(\phi_1,\ldots, \phi_n)^t, \ \ \phi_j = - \psi_0(z_j), \ \ j=1,\ldots,n.
\eeq
\end{proposition}
Proposition~1 is a variation of statements used in the book \cite{AGHH}.

\section{Zero-energy bound states}
\begin{theorem}
Equation (\ref{eq:1}) with multipoint potential $v$ of the form (\ref{eq:2}) admits a non-zero solution $\psi\in L^2(\RR^3)$
at energy $E=0$ if and only if there exists a non-zero $\vec q$ such that 
\beq
\label{eq:12}
A\vec q=0 \ \ \mbox{for} \ \  E=0,
\eeq
\beq
\label{eq:13}
\sum\limits_{j=1}^n q_j=0,
\eeq
where $A$ is defined by (\ref{eq:10}). In addition, the one-to-one correspondence between such solutions $\psi$ and 
vectors $\vec q$ is given by:
\beq
\label{eq:14}
\psi(x) = -\frac{1}{4\pi }\sum\limits_{j=1}^n q_j \frac{1}{|x-z_j|}.
\eeq
\end{theorem}
Theorem~1 follows from Proposition~1, the property that 
$$
G^+(|\cdot|,0)\in L^2_{\mathrm{loc}}(\RR^3),
$$ 
the linear independence of $G^+(|\cdot-z_j|,0)$, $j=1,\ldots,n$,
the following asymptotic formula for $\psi$ of the form (\ref{eq:14}):
\beq
\label{eq:15}
\psi(x) =  -\frac{1}{4\pi |x|}\sum\limits_{j=1}^n q_j+ O\left(\frac{1}{|x|^2}\right) \ \ \mbox{as} \ \ 
|x|\rightarrow+\infty,
\eeq
and the following lemma:
\begin{lemma}
Let $\psi_0$ satisfy (\ref{eq:7}) for $E=0$ and: 
\beq
\label{eq:15.2}
\psi_0=\psi_{0,1}+\psi_{0,2}, \ \ \psi_{0,1}(x)=o(1) \ \ \mbox{as} \ \ 
|x|\rightarrow\infty, \ \ \psi_{0,2}\in L^2(\RR^3).
\eeq 
Then $\psi_0\equiv0$.  
\end{lemma}
Lemma~1 follows from the mean value property over balls for harmonic functions, the Cauchy-Schwarz inequality and 
Liouville's theorem for harmonic functions.

In the next example we consider a scatterer consisting of four equal single point scatterers located in the vertices of 
a regular tetrahedron.

\begin{example} 
Let $n=4$, $z_j\in\RR^3$, $|z_j-z_{j'}|=s>0$ for all $j\ne j'$, $1\le j,j'\le 4$,
$\alpha_j=\alpha=-(4\pi s)^{-1}$, and $v$ be given by (\ref{eq:2}).
Then $E=0$ is a triple eigenvalue for equation~(\ref{eq:1}) (i.e. the space of solutions of equation (\ref{eq:1}) in 
$L^2(\RR^3)$ for $E=0$ is three-dimensional).  
\end{example}
This statement follows directly from Theorem~1.

In the next example we consider a scatterer consisting of $2m$ equal single point scatterers located in the vertices of 
a regular planar $2m$-gon.
\begin{example}
Let $n=2m$, $z_1,\ldots,z_{2m}\in\RR^3$ be sequentially enumerated vertices of a convex regular planar (belonging to a fixed plane) polygon with $2m$ vertices,  
\beq
\label{eq:16}
\alpha_j=\alpha=-\sum\limits_{k=2}^{2m}\frac{(-1)^k}{4\pi |z_k-z_1|},
\eeq
and $v$ be given by (\ref{eq:2}).
Then: 
\beq
\label{eq:17}
\alpha\ne 0,
\eeq
\item 
\beq
\label{eq:18}
\psi(x) = -\frac{1}{4\pi }\sum\limits_{j=1}^{2m}  \frac{(-1)^{j+1}}{|x-z_j|}
\eeq
is a zero-energy bound state for equation ~(\ref{eq:1});
\beq
\label{eq:19}
\psi(x) = O\left(\frac{1}{|x|^{m+1}}  \right) \ \ \mbox{as} \ \ |x|\rightarrow+\infty.
\eeq
\end{example}
The point is that in this example the zero-energy bound state $\psi$ is strongly localized for large $m$. 

On the other hand, this example holds already for $m=1$ when our polygon reduces to a segment. 

In addition, we have the conjecture that $E=0$ is a simple eigenvalue in this example; it was checked numerically 
up to $m=48$ using Theorem~1.

The property (\ref{eq:17}) follows from the formulas: 
\beq
\label{eq:20}
\alpha=-\sum\limits_{k=2}^{m+1} (-1)^k u_k, \ \ 
u_k=\left\{
\begin{array}{ll}
\frac{1}{2\pi|z_k-z_1|}, & k=2,\ldots,m,\\
\frac{1}{4\pi|z_{m+1}-z_1|}, & k=m+1,  
\end{array} 
\right. 
\eeq
\beq
\label{eq:21}
u_2>u_3>\ldots> u_{m+1}>0.
\eeq

Formulas (\ref{eq:16}), (\ref{eq:18}) were obtained using (\ref{eq:12}) with $q_j=(-1)^{j+1}$, $j=1,\ldots,n$, and finding
$\alpha$ such that (\ref{eq:12}) holds for $\alpha_j=\alpha$, $j=1,\ldots,n$.

To prove the localization property (\ref{eq:19}) we choose orthogonal coordinates such that 
\begin{align}
z_j=r_0 \,\omega_j, \ \ r_0>0,\ \  \omega_j=\left(\cos\left(\frac{\pi (j-1)}{m}\right), \sin\left(\frac{\pi (j-1)}{m} \right),0 \right),\nonumber \\ 
j=1,\ldots,2m.\label{eq:22}
\end{align}
We have 
\beq
\label{eq:23}
\frac{1}{|x-z_j|} = \frac{1}{\left(R^2+r_0^2\right)^{1/2}} \sum\limits_{l=0}^{+\infty} b_l 
\left(\frac{2r_0 R}{R^2+r_0^2} \right)^l (\nu\omega_j)^l, \ \ R\rightarrow\infty,
\eeq
where
$$
R=|x|, \ \  \nu= x/|x|, \ \ \nu=(\sin\theta \cos\phi, \sin\theta \sin\phi, \cos\theta),
$$ 
$\theta$, $\phi$ are the polar and azimuthal angles of $\nu$, respectively, $b_l$ are the expansion coefficients:
\beq
\label{eq:24}
(1-t)^{-1/2}=\sum\limits_{l=0}^{+\infty} b_l t^l, \ \ |t|<1.
\eeq
Thus,
\begin{align}
\label{eq:25}
&\psi(x)=\frac{1}{4\pi}\frac{1}{\left(R^2+r_0^2\right)^{1/2}} \sum\limits_{l=0}^{+\infty} b_l 
\left(\frac{2r_0 R}{R^2+r_0^2} \right)^l\left[ \sum\limits_{j=1}^{2m} (-1)^j (\nu\omega_j)^l  \right], \ \ R\rightarrow\infty,\\
&  \nu\omega_j = \sin\theta \cos\left(\phi-\frac{\pi(j-1)}{m}  \right).\nonumber
\end{align}
The localization  (\ref{eq:19}) follows from the property:
\beq
\label{eq:26}
{\cal C}_l := \sum\limits_{j=1}^{2m} (-1)^j (\nu\omega_j)^l =0 \ \ \mbox{for} \ \ 0\le l\le m-1.
\eeq
In turn, identity  (\ref{eq:26}) follows from the formulas:
\beq
\label{eq:27}
{\cal C}_l={\cal C}_l(\theta,\phi)=(\sin\theta)^l \ \ \sum\limits_{k=-l}^l c_{lk} e^{ik\phi} \ \ 
\mbox{for some} \ \ c_{lk} \ \ \mbox{depending on} \ \ m;
\eeq
\beq
\label{eq:28}
{\cal C}_l(\theta,\phi+\pi/m)=-{\cal C}_l(\theta,\phi).
\eeq
This completes the proof of Example~2.

\end{document}